\documentstyle[12pt]{article}

\def\lag{{\cal L}}
\def\Aslash{A\hbox to 0mm{\kern-2.3mm/\hss}}
\def\Dslash{D\hbox to 0mm{\kern-2.75mm/\hss}}
\def\dslash{\partial\hbox to 0mm{\kern-2.2mm/\hss}}
\def\qslash{q\hbox to 0mm{\kern-1.75mm/\hss}}

\begin{document}

\begin{titlepage}

\def\mytoday#1{{ } \ifcase\month \or
 January\or February\or March\or April\or May\or June\or
 July\or August\or September\or October\or November\or December\fi
 \space \number\year}

\vspace*{-2.15cm}
\noindent\hbox to\textwidth{\hfill hep-ph/9610279}
\noindent\hbox to\textwidth{\hfill TTP96-47}
\vspace*{.3cm}
\begin{center}
{\LARGE A Manifestly Gauge-Invariant Approach to Charged Particles}
\footnote{Work supported in part by DFG Grant Nr.\ Ma1187/7-2.} \\ [1.0cm]


   A. Schenk

\vspace*{.9cm}
    Universit\"at Karlsruhe\\
    Institut f\"ur Theoretische Teilchenphysik\\
    Kaiserstr.\ 12\\
    D-76128 Karlsruhe, Germany \\[.9cm]

 September 1996
\vspace*{.5cm}

\begin{abstract}
In this article we provide a manifestly gauge-invariant approach to
charged particles.  It involves (1) Green functions of
gauge-invariant operators and (2) Feynman rules which do not depend on
any kind of gauge-fixing condition. First, we provide a thorough
analysis of QED. We propose a specific set of gauge-invariant Green
functions and describe a manifestly gauge-invariant technique to
evaluate them.  Furthermore we show how Green's functions and Feynman
rules of the manifestly gauge-invariant approach and of the 
Faddeev-Popov ansatz are related to each other.  Finally, we extend
this manifestly gauge-invariant approach to the Standard Model of
electroweak interactions. 

Motivation for such an approach is abundant.  A gauge-dependent
framework does obstruct not only theoretical insight but also
phenomenological analyses of precision experiments at LEP.
Unresolved issues include the analysis of finite width effects of
unstable particles and of oblique corrections parametrised by, e.g.,
$S, T,$ and~$U$. Quite another example of considerable complications
caused by a gauge-dependent framework is the matching of a full and an
effective theory, relevant for effective field theory analyses of the
symmetry-breaking sector of the Standard Model.
\end{abstract}
\end{center}
\end{titlepage}

\section{Introduction}

Almost every analysis of gauge theories involves a manifestly
gauge-dependent technique based on gauge-dependent Green functions.
This gauge-depend\-ence, however, manifests itself only in the off-shell
behaviour of these functions. Their pole positions and residues, i.e.,
particle masses, decay widths, and $S$-matrix elements, are
gauge-independent. Due to this property such functions are interesting
and useful objects for physicists.

Naively this appears to be all there is to say. As long as the results
are gauge-independent there seems to be no reason at all why the
symmetry should be manifest throughout the whole calculation. The
great success with which gauge-dependent techniques are employed seems
indeed to support this notion.

Real life, however, is a more complicated matter. The literature
contains quite a number of examples~\cite{Cerny,DKS,glue,DDW,MGDGF,EsMa} 
where the gauge-dependence of the perturbative approach does obstruct
not only theoretical insight but also phenomenological analyses of
precision experiments at presently accessible energies.

Several attempts~\cite{Cerny, DKS, glue, EsMa, pinch, BGFM} have
recently been made in order to resolve these issues. The techniques
employed differ in the degree with which symmetry properties are
manifest. The important point, however, is that they all are still
gauge-dependent.

The purpose of this article is to provide a manifestly gauge-invariant
approach to analyse gauge theories with charged particles. It does not
involve any kind of gauge-dependence. Its two main ingredients are (1)
Green functions of gauge-invariant operators and (2) Feynman rules
which do not involve any kind of gauge fixing. Some aspects of this
approach have already been discussed in the
literature~\cite{AHM,Corrigan_Goddard} where it has been applied to
specific problems. Some important issues are, however, still
unresolved. They include the particular treatment of charged particles
which we will focus on in this work. Before going into detail, we
would like to recall what kind of problems are encountered in a
gauge-dependent approach and why this is relevant to phenomenological
analyses.

Oblique corrections describe the effects of new physics on the vacuum
polarisation of the Standard Model gauge-bosons currently measured
with precision electroweak experiments~\cite{PDG}.  Since these
corrections involve off-shell properties of the corresponding Green
functions it is by no means clear how to provide a physical, i.e.,
gauge-independent definition of these quantities within the usual
gauge-dependent framework. As a matter of fact, bosonic contributions
to the relevant self-energies turn out to be gauge-dependent already
at the one-loop level~\cite{DKS}. On the other hand, any quantity
which is intended to parametrise new physics' effects of a theory yet
unknown should obviously be free of any gauge artefacts.

Problems of this kind arise whenever certain gauge-dependent
contributions cannot cancel each other order by order in perturbation
theory.  Other examples include the analyses of finite-width effects
for unstable particles~\cite{Cerny} or of a finite gluon mass~\cite{glue}
which involve the re-summation of certain subsets of Feynman diagrams.
As a result physical quantities turn out to depend on the gauge-fixing
parameter.

One attempt to remove this problem is the so called pinch
technique~\cite{glue, pinch}. It sticks to the inherently gauge-dependent
framework, yet adds a number of rules about how to modify 1PI
functions such that the gauge-parameter dependence of physical
quantities disappears.

The background field method~\cite{BGFM} is another attempt. Through a
particular choice of gauge-fixing condition the background field
action turns out to be invariant under gauge transformations of the
background fields. This property of the effective action may indeed be
quite useful in certain applications. Nevertheless, the background
field method deals with gauge-dependent Green functions.

Both approaches fail to resolve the problems described above. This may
be illustrated with the attempts to provide a gauge-independent
parametrisation of oblique corrections. It was mentioned above that
bosonic contributions to the relevant self-energies are
gauge-dependent if perturbation theory is applied in the
straightforward fashion. Both, the pinch technique and the background
field method were used to further investigate this problem.

The attempt based on the pinch technique yields a result for
bosonic one-loop corrections which does not depend on the gauge-fixing
parameter~\cite{DKS}. The approach based on the background field method,
on the other hand, yields corrections which still depend on the
gauge-fixing parameter~\cite{DDW}. For the particular choice~$\xi=1$ both
results agree.

First of all, this reminds us of the fact that independence of the
gauge-fixing parameter is only a necessary condition for
gauge-independence, not a sufficient one. Moreover, it demonstrates
that this handle does not help to pin down the physical contents of
oblique corrections, analysed within a gauge-dependent framework. At
present this must still be considered an unresolved issue. 

Due to the finite width of the $W$-boson gauge invariance is also an
issue for the analysis of $W$-pair production at LEP. A detailed account
of the problems of various gauge-dependent approaches can be found in
the report of the LEP working group~\cite{Cerny}.

Effective field theories provide another example of unwanted
complications caused by a gauge-dependent framework. The low-energy
structure of a theory containing light and heavy particle species
which are separated by a mass gap can adequately be described by an
effective theory which contains only the light particles. The Standard
Model with a heavy Higgs boson is an example which has recently
received considerable attention~\cite{heavy_Higgs}. In order to
determine the effective Lagrangian, which describes the effective
field theory, one requires that corresponding Green functions in both
theories have the same low-energy structure. If a gauge-dependent
framework is used~\cite{MGDGF,EsMa} one again has to make sure that no
gauge artefacts enter the low-energy constants of the effective
Lagrangian.  A detailed discussion of the complications involved with
matching gauge-dependent Green functions may be found in
Ref.~\cite{AHM}.

To avoid these problems one should match only gauge-invariant
quantities, such as $S$-matrix elements~\cite{EsMa}. As it turns out,
however, matching $S$-matrix elements is quite cumbersome.  Functional
techniques~\cite{LSM}, which involve Green's functions, are much
easier to use. A manifestly gauge-invariant technique combines both
advantages: The absence of gauge-artefacts and the simplicity and
elegance of the functional approach~\cite{AHM}.

There certainly is ample motivation to provide a manifestly
gauge-invariant framework to analyse gauge theories. This work focuses
on the particulars of charged particles. In addition to formulating this
approach we also compare it thoroughly with the usual approach based
on the Faddeev-Popov ansatz.

The present article is organised as follows: In the next section we
provide a thorough analysis of a manifestly gauge-invariant approach
to QED. In particular, we propose a specific set of gauge-invariant
Green functions and describe a manifestly gauge-invariant technique
to evaluate them. Feynman rules are summarised at the end of the
section. Furthermore we show how Green's functions and Feynman rules
of this approach and of the Faddeev-Popov ansatz are related to each
other. Section 3 contains a detailed discussion of the results. In
Section 4 we extend this manifestly gauge-invariant approach to the
Standard Model of electroweak interactions. Again we compare the Green
functions of this approach with those of the Faddeev-Popov ansatz.
Finally, we summarise our results in Section 5.

\section{QED}

We consider Green's functions of the fermion field~$\psi$ and the
electromagnetic field strength $F_{\mu\nu}$. If there were no gauge
symmetry these Green functions would be generated by the
vacuum-to-vacuum transition amplitude
\begin{equation} \label{ill_func}
e^{i W[\eta, \bar\eta, k_{\mu\nu}]} = 
\langle 0_{\rm out}|0_{\rm in}\rangle_{\eta,\bar\eta, k_{\mu\nu}} 
=\langle0| T\left[ e^{i \bar\eta \psi_{\rm op} + i \bar\psi_{\rm op}\eta
+ i k_{\mu\nu} F^{\mu\nu}_{\rm op}} \right]|0\rangle \ .
\end{equation}
However, this amplitude is associated with the Lagrangian
\begin{eqnarray}
\lag &=& \lag_{QED} +  \bar\eta \psi + \bar\psi\eta
+  k_{\mu\nu} F^{\mu\nu} \\
\lag_{QED} &=& \bar\psi\left( i \Dslash - m\right) \psi - {1\over 4 e^2}
F_{\mu\nu} F^{\mu\nu}   \label{lag_qed}
\end{eqnarray}
whose source-independent piece~$\lag_{QED}$ is invariant under
gauge transformations of the form
\begin{eqnarray}
  \psi &\rightarrow& e^{- i \omega}\psi \ , \\
  A_\mu &\rightarrow& A_\mu + \partial_\mu \omega \ .
\end{eqnarray}
The covariant derivative is given by
\begin{equation}
D_\mu = \partial_\mu + i A_\mu \ .
\end{equation}

The external source~$\eta$, on the other hand, is a fixed function of
space-time. Hence, Green's functions involving the fermion
fields~$\psi$ and~$\bar\psi$ do not reflect the symmetry properties of
the Lagrangian~$\lag_{QED}$.  What is worse, the generating functional given
in Eq.~(\ref{ill_func}) is not even well-defined.  The general problem
associated with gauge symmetries is readily worked out in the
path-integral
representation of the generating functional, which is of the form
\begin{equation} \label{ppathrep}
e^{i W[\eta, \bar\eta, k_{\mu\nu}]} = \int d\mu[\psi,\bar\psi,A_\mu]
e^{i \int {\rm d}^{\rm d}x\, \lag_{QED} 
+ \bar\eta \psi + \bar\psi\eta +  k_{\mu\nu} F^{\mu\nu}}  \ .
\end{equation}

At leading order in the loop expansion the generating functional is
given by the classical action
\begin{equation}
W[\eta, \bar\eta, k_{\mu\nu}] = \int {\rm d}^{\rm d}x\,
\lag_{QED}(\psi^{{\rm cl}}, A_\mu^{{\rm cl}}) 
+ \bar\eta \psi^{{\rm cl}} + \bar\psi^{{\rm cl}}\eta +  k^{\mu\nu} F_{\mu\nu}^{{\rm cl}}
\ .
\end{equation}
The classical fields $A_\mu^{{\rm cl}}$ and $\psi^{{\rm cl}}$ are determined by the
condition that the phase of the integrand in Eq.~(\ref{ppathrep}) be 
stationary, i.e., that
\begin{equation} \label{stationary}
\delta \int {\rm d}^{\rm d}x\, 
\lag_{QED} + \bar\eta \psi + \bar\psi \eta +  k_{\mu\nu} F^{\mu\nu} =
0 \ .
\end{equation}
Since only the source terms involving the external field~$\eta$
violate the gauge symmetry, variations of the fields corresponding to
infinitesimal gauge transformations yield the condition
\begin{equation} \label{condition}
\bar\eta\psi^{{\rm cl}} - \bar\psi^{{\rm cl}}\eta = 0 \ .
\end{equation}
It is a concise expression for the problem of the amplitude in
Eq.~(\ref{ill_func}). On the one hand, gauge invariance wants the
photon to couple to conserved currents only. This requires
condition~(\ref{condition}) to be satisfied. The external
source~$\eta$, on the other hand, does not want to do us this
favour. The solutions of the equations of motion do not satisfy this
condition. It is, in fact, obvious that the electromagnetic current
cannot be conserved in the presence of sources which create or destroy
charged particles.  In other words, the generating functional~$W[\eta,
\bar\eta, k_{\mu\nu}]$ is not properly defined.

The usual remedy is to abandon gauge invariance. A supplementary
condition of the form
\begin{equation} \label{gaugecondition}
f(A_\mu) = 0 
\end{equation}
is introduced, which freezes the gauge degree of freedom.  Green's
functions are then defined through the Faddeev-Popov
ansatz~\cite{FadPop} for the generating functional, i.e., through
\begin{equation} \label{fadpopans}
e^{i W_f[\eta, \bar\eta, k_{\mu\nu}]} \doteq \int d\mu[\psi,\bar\psi,A_\mu]
 \det(M_f) \delta\left(f(A_\mu)\right) e^{i \int {\rm d}^{\rm d}x\, \lag_{QED} 
+ \bar\eta \psi + \bar\psi\eta +  k_{\mu\nu} F^{\mu\nu}}  \ .
\end{equation}
The matrix~$M_f$ is the variation of the functional~$f$ with respect
to a gauge transformation. This ansatz defines gauge-dependent
Green functions.

There is, however, another solution to the problem. Note that the
source term involving the external field~$k_{\mu\nu}$ does not affect
condition~(\ref{condition}). This follows from the gauge invariance of
the field strength. Hence one may also define Green's functions
through the vacuum-to-vacuum transition amplitude
\begin{equation} \label{vacvac}
e^{i \bar W[N, \bar N, k_{\mu\nu}]} = 
\langle 0_{\rm out}|0_{\rm in}\rangle_{N,\bar N, k_{\mu\nu}} 
=\langle0| T\left[ e^{i \bar N \psi_{\rm op} + i \bar\psi_{\rm op} N
+ i k_{\mu\nu} F^{\mu\nu}_{\rm op}} \right]|0\rangle 
\end{equation}
which involves a gauge-variant external source~$N$ that transforms
according to
\begin{equation} \label{gis}
N \rightarrow e^{-i\omega} N
\end{equation}
under gauge transformations. Sources of this form do not couple to the
gauge degree of freedom and, thus, define gauge-invariant Green
functions.  Problem~(\ref{condition}) does not appear. We will see
below how this comes about.

From a physical point of view the second approach makes even more
sense. After all, gauge invariance means that the phase of the fermion
field is completely arbitrary at any space-time point -- and
physically unobservable. Hence, any artificial dependence on the gauge
degree of freedom should be expected to do more harm than good.

The path-integral representation of the amplitude~(\ref{vacvac}) is
of the form
\begin{equation} \label{pathreprep}
e^{i \bar W[N, \bar N, k_{\mu\nu}]} = \int d\mu[\psi,\bar\psi,A_\mu]
e^{i \int {\rm d}^{\rm d}x\, \lag_{QED} 
+ \bar N \psi + \bar\psi N +  k_{\mu\nu} F^{\mu\nu}}  \ .
\end{equation}
Note that no gauge-fixing condition is introduced, in contrast to the
ansatz in Eq.~(\ref{fadpopans}). In the next subsection we will
evaluate this functional for a specific external source~$N$.

Beforehand, we should add some remarks about gauge fixing. The
vacuum-to-vacuum transition amplitude~(\ref{ill_func}) and the
corresponding path-integral representation~(\ref{ppathrep}) are not
well-defined because the external source~$\eta$ couples to the gauge
degree of freedom. In order to define Green's functions properly in the
presence of such sources one has to fix the gauge and work with the
Faddeev-Popov ansatz~(\ref{fadpopans}). 

The situation is quite different if only gauge-invariant sources are
considered. In this case one may also fix the gauge in order to
evaluate the path integral. However, one need not do so. Both ways
yield the same Green functions. For example, the generating
functional~$\bar W[0,0,k_{\mu\nu}]$ defined in Eq.~(\ref{pathreprep})
is exactly the same as the generating functional~$W_f[0,0,k_{\mu\nu}]$
defined in Eq.~(\ref{fadpopans}). The difference is only seen at
intermediate steps of a calculation. If the gauge is fixed,
contributions from certain diagrams depend on the gauge-fixing
parameter~$\xi$. In contrast to the Faddeev-Popov ansatz with
gauge-dependent sources, however, this dependence drops out order by
order in perturbation theory if all contributions to a given Green
function are combined -- even off-shell.

Thus, if one is interested in a specific subset of contributions only,
like, for instance, oblique corrections, it is better to evaluate the
generating functional without gauge fixing. Otherwise one will still
encounter the complications one actually wants to avoid, even if
gauge-invariant Green functions are considered. This is also the case
for the analysis of finite width effects of unstable particles.

Note that this remarks do not only apply to the specific case we are
considering here, but are valid in any gauge theory, including the
Standard Model of electroweak interactions and QCD.

\subsection{QED the manifestly gauge-invariant way}

It is not difficult to see that sources with transformation properties as
specified in Eq.~(\ref{gis}) do exist. The most straightforward
construction may be of the form
\begin{equation} \label{fstinvsource}
N(x) = \eta(x) e^{ -i\int_{-\infty}^x {\rm d}x_\mu^\prime
A^\mu(x^\prime)} \ ,
\end{equation}
which can already be found in Ref.~\cite{tHooft}. As long as one is
only interested in the existence of such sources and some of their
formal properties, this choice will surely suffice.  However, the
corresponding Green functions have unwanted properties. In particular,
they depend on the specific paths used to evaluate the integrals in
Eq.~(\ref{fstinvsource}) for each external leg. Furthermore, this
source depends on the dynamical component of the gauge field.

Another source term with the same transformation properties is of the
form
\begin{equation} \label{sndinvsource}
N(x) = \eta(x) e^{ - i \int {\rm d}^{\rm d}y G_0(x-y)
\partial_\mu A^\mu(y) } \ ,
\end{equation}
where the propagator~$G_0$ is defined as
\begin{equation}\label{masslessprop}
        \Box G_0(z) = \delta^{(d)}(z) \ .
\end{equation}
It does not suffer from the path-dependence the other source has. And
it does not involve any dynamical degree of freedom. This is exactly
the source whose properties we will investigate here.  Thus, from now
on, the external field~$N$ is of the particular form given in
Eq.~(\ref{sndinvsource}).

We define Green's functions involving the fermion field through
derivatives of the generating functional~(\ref{pathreprep}) with
respect to the external field~$\eta$. They differ by a phase from
corresponding derivatives with respect to the external source~$N$. To
make this definition more explicit we redefine the generating
functional as
\begin{equation} \label{pathrep}
e^{i W[\eta, \bar\eta, k_{\mu\nu}]} \doteq \int d\mu[\psi,\bar\psi,A_\mu]
e^{i \int {\rm d}^{\rm d}x\, \lag_{QED} 
+ \bar N \psi + \bar\psi N +  k_{\mu\nu} F^{\mu\nu}}  
\end{equation}
for the source~$N$ given in Eq.~(\ref{sndinvsource}).

It is not quite straightforward to write down Feynman rules to
evaluate the functional given in Eq.~(\ref{pathrep}). First of all, we
do not want to introduce any gauge-fixing condition in the path
integral for reasons described above. Hence, the usual definition of
the gauge-field propagator is of no avail. Furthermore, the source~$N$
in Eq.~(\ref{sndinvsource}) involves an exponential factor which, one
would think, should produce vertices that emit an arbitrary number of
photons. This presumption, however, will turn out to be wrong.

Thus, in order to provide a sound understanding of our manifestly
gauge-invariant approach to Green's functions, we will go step by step
through the loop expansion. Feynman rules, which we think will turn
out to be much simpler than anticipated, are given at the end of this
section.

\subsubsection*{Trees}

At leading order in the loop expansion the generating
functional~(\ref{pathrep}) is given by the classical action
\begin{equation}
W[\eta, \bar\eta, k_{\mu\nu}] = \int {\rm d}^{\rm d}x\,
\lag_{QED}(A^{{\rm cl}}_\mu, \psi^{{\rm cl}})
+ \bar N \psi^{{\rm cl}} + \bar\psi^{{\rm cl}} N +  k^{\mu\nu} F_{\mu\nu}^{{\rm cl}} 
\end{equation}
where the fields~$A_\mu^{{\rm cl}}$ and~$\psi^{{\rm cl}}$ satisfy the equations of
motion
\begin{eqnarray}
\left(i\Dslash - m\right)\psi^{{\rm cl}} &=& -N \ , \label{Neqmo}\\
\bar\psi^{{\rm cl}}\left(i\stackrel{\leftarrow}{\Dslash} - m\right) &=& -\bar N \ ,\label{Nbeqmo} 
\end{eqnarray}
and
\begin{eqnarray}
\partial^\mu F_{\mu\nu}^{{\rm cl}}  &=& 2 e^2 \partial^\mu k_{\mu\nu} +
 e^2 \bar\psi^{{\rm cl}}\gamma_\nu\psi^{{\rm cl}} 
+ i e^2 \partial_\nu {1\over\Box} \left(\bar N\psi^{{\rm cl}} - \bar\psi^{{\rm cl}} N\right)
\label{Aeqmo}\ ,
\end{eqnarray}
with
\begin{equation}
\stackrel{\leftarrow}{D}_\mu = - \stackrel{\leftarrow}{\partial}_\mu + i A_\mu^{{\rm cl}} \ .
\end{equation}
The third term on the right hand side of Eq.~(\ref{Aeqmo}) is due to
the variation of the phase in Eq.~(\ref{sndinvsource}). Its effect is
readily worked out. 
Eqs.~(\ref{Neqmo}) and~(\ref{Nbeqmo}) imply the identity
\begin{equation} \label{divcurrent}
\partial_\mu\left(\bar\psi^{{\rm cl}}\gamma^\mu\psi^{{\rm cl}}\right) = i \left(\bar\psi^{{\rm cl}} N -
\bar N \psi^{{\rm cl}}\right) \ .
\end{equation}
It shows that the electromagnetic current is not conserved in the
presence of an external source which emits or absorbs charged
particles (cf.~problem~(\ref{condition})).  Gauge invariance, on the other hand, requires that the
photon field couples to conserved currents only. The third term on the
right hand side of Eq.~(\ref{Aeqmo}) ensures that this is indeed the
case.  Using Eq.~(\ref{divcurrent}) one obtains
\begin{equation}
\partial^\mu F_{\mu\nu}^{{\rm cl}}  = 2 e^2  \partial^\mu k_{\mu\nu} +
 e^2 {\rm PT}_{\nu\rho} \bar\psi^{{\rm cl}}\gamma^\rho\psi^{{\rm cl}} 
\label{AAeqmo}\ ,
\end{equation}
where
\begin{equation}
{\rm PT}_{\mu\nu} = g_{\mu\nu} - {\rm PL}_{\mu\nu} \quad{\rm and}\quad
{\rm PL}_{\mu\nu} = \partial_\mu {1\over\Box} \partial_\nu
\end{equation}
project onto the transversal and longitudinal degrees of freedom
respectively. If the external source~$\eta$ is switched off the
electromagnetic current is conserved and the transversal projector in
Eq.~(\ref{AAeqmo}) reduces to the identity.

Gauge invariance implies that the equations of motion have a whole
class of solutions. Every two representatives are related to each
other by a gauge transformation. In order to solve these equations
we write the fermion field as (cf.~Eq.~(\ref{sndinvsource}))
\begin{equation} \label{rescalepsi}
\psi^{{\rm cl}}(x) = \chi(x) e^{ - i \int {\rm d}^{\rm d}y G_0(x-y)
\partial^\mu A^{{\rm cl}}_\mu(y) } \ . 
\end{equation}
As a result, the phase drops out of Eq.~(\ref{Neqmo}), which takes
the form
\begin{equation}
\left(i\Dslash^T - m\right)\chi = -\eta \ .
\label{Nheqmo}
\end{equation}
It involves only the transversal component $A_\mu^T \doteq {\rm
PT}_{\mu\nu} A^\nu$ of the gauge field through the derivative
\begin{equation}
        D^T_\mu \doteq \partial_\mu + i A_\mu^T \ .
\end{equation}
The longitudinal component
$A_\mu^L \doteq {\rm PL}_{\mu\nu} A^\nu$ does not enter
Eq.~(\ref{AAeqmo}) either. Hence, the solutions of the equations of
motion are of the form
\begin{eqnarray}
A^{L, {\rm cl}}_\mu & = & \partial_\mu\omega  \label{Alsol} \ ,\\
\psi^{{\rm cl}} & = & e^{-i\omega} \chi  \label{psisol} \ , \\
A^{T,{\rm cl}}_\mu(x) &=& i e^2  \int  {\rm d}^{\rm d}y \Delta_{\mu\nu}(x-y)\left(
  2 \partial_\rho k^{\rho\nu}(y) + (\bar\chi\gamma^\nu\chi)(y)
\right)  \label{atsol} \ , \\
\chi(x) &=& i \int  {\rm d}^{\rm d}y S(x-y) \left( -\eta(y) +
\left(\Aslash^{T,{\rm cl}}\chi\right)(y)\right) \ , \label{chisol}
\end{eqnarray}
with an undetermined function~$\omega$ describing the gauge degree of
freedom. It is important to note that this function does not enter
the generating functional in Eq.~(\ref{pathrep}). Gauge invariance
ensures that it drops out. Hence, Green's functions do not depend on
this function either. This justifies the redefinition of the
generating functional in Eq.~(\ref{pathrep}). The free propagators 
are defined as
\begin{eqnarray} \label{GBpropagator}
\Delta_{\mu\nu}(x-y) &=& i \langle x|{{\rm
PT}_{\mu\nu}\over -\Box}|y\rangle \ ,\\
S(x-y) &=& i\langle x|{-1 \over i \dslash - m}|y\rangle \ . \label{Spropagator}
\end{eqnarray}
Gauge invariance is indeed manifest.

\subsubsection*{One Loop}

A convenient way to evaluate the one-loop contribution to the
generating functional~(\ref{pathrep}) is the method of steepest
descent. Using the parametrisation
\begin{eqnarray} \label{Afstfluc}
A_\mu &=& A_\mu^{{\rm cl}} + \sqrt{2} e q_\mu \\
\psi &=& \psi^{{\rm cl}} + \kappa \label{psifstfluc}
\end{eqnarray}
for the quantum fluctuations one obtains for the one-loop approximation
\begin{eqnarray} \label{fstaction}
\lefteqn{e^{i W[\eta, \bar\eta, k_{\mu\nu}]} }\\
& = &e^{i \int {\rm d}^{\rm d}x\, \lag(A_\mu^{{\rm cl}}, \psi^{{\rm cl}})}
\int d\mu[\kappa,\bar\kappa,q_\mu]
 e^{i \int {\rm d}^{\rm d}x\, 
\bar\kappa D_{\bar\kappa\kappa}\kappa +
\bar\kappa D_{\bar\kappa q}^\nu q_\nu +
q_\mu D_{q\kappa}^{\mu}\kappa +
q_\mu D_{qq}^{\mu\nu} q_\nu }  \nonumber 
\ , \nonumber 
\end{eqnarray}
where
\begin{eqnarray}
\lag & = & \lag_{QED} + \bar N \psi + \bar\psi N +  k_{\mu\nu} F^{\mu\nu}
\ , \\
D_{\bar\kappa\kappa} & = & i\Dslash - m \ , \label{dkk}\\
D_{\bar\kappa q}^\nu         & = &  -\sqrt{2} e \left(\gamma^\nu \psi^{{\rm cl}} + i N
                {1\over\Box}\partial^\nu\right) \ , \label{dkq} \\
D_{q\kappa}^{\mu}       & = & -\sqrt{2} e \left(\bar\psi^{{\rm cl}}\gamma^\mu 
                + i \partial^\mu {1\over\Box}\bar N\right) \ , \label{dqk} \\ 
D_{qq}^{\mu\nu}           & = & \Box g^{\mu\nu} - \partial^\mu\partial^\nu
   + e^2 \partial^\mu {1\over\Box} \left( \bar N \psi^{{\rm cl}} + \bar \psi^{{\rm cl}}
        N\right) {1\over\Box}\partial^\nu \ . \label{dqq}
\end{eqnarray}
All terms involving the free massless propagator~$\Box^{-1}$ again arise from the 
variation of the phase in Eq.~(\ref{sndinvsource}). We proceed by
diagonalizing the quadratic form in the exponent of
Eq.~(\ref{fstaction}) with the shift
\begin{equation}
\kappa \rightarrow \kappa + \sqrt{2} e (i\Dslash-m)^{-1}
\left( \qslash\psi^{{\rm cl}} + i N {1\over\Box} \partial_\mu q^\mu\right) 
\end{equation}
and obtain the diagonal matrix
\begin{eqnarray} 
D_{\bar\kappa\kappa} & = & i\Dslash - m \ , \\
D_{qq}^{\mu\nu}           & = & \Box g^{\mu\nu} - \partial^\mu\partial^\nu
   + e^2 \partial^\mu {1\over\Box} \left( \bar N \psi^{{\rm cl}} + \bar \psi^{{\rm cl}}
        N\right) {1\over\Box}\partial^\nu \label{diagonaldqq} \\ \nonumber
& & \quad\mbox{} - 2 e^2 
\left( \bar\psi^{{\rm cl}} \gamma^\mu + i \partial^\mu{1\over\Box} \bar N\right)
(i\Dslash-m)^{-1}
\left( \gamma^\nu\psi^{{\rm cl}} + i N \partial^\nu{1\over\Box} \right) \ .
\end{eqnarray}
Next we employ the equations of motion~(\ref{Neqmo})
and~(\ref{Nbeqmo}) to remove all occurrences of the external
fields~$N$ and~$\bar N$. One obtains, for example, 
\begin{equation}
 \gamma^\nu\psi^{{\rm cl}} + i N \partial^\nu{1\over\Box} =
\gamma_\mu\psi^{{\rm cl}} {\rm PT}^{\mu\nu}
- i (i\Dslash-m) \psi^{{\rm cl}} \partial^\nu {1\over\Box} \ .
\end{equation}
The final result turns out to be
\begin{equation} \label{fullAprop}
D_{qq}^{\mu\nu}   =  \Box g^{\mu\nu} -
\partial^\mu\partial^\nu - {\rm PT}^{\mu\alpha} \sigma_{\alpha\beta}  {\rm PT}^{\beta\nu}
\end{equation}
with
\begin{equation} \label{sigmadef}
\sigma_{\mu\nu} = 2 e^2\bar\psi^{{\rm cl}} \gamma_\mu
(i\Dslash-m)^{-1} \gamma_\nu\psi^{{\rm cl}} \ .
\end{equation}
It is quite remarkable. All contributions from the variation of the
source term~$\bar N\psi + \bar\psi N$, which seemed prone to create
additional infrared singularities, transformed into
the apparent transversal structure of the full
propagator~(\ref{fullAprop}). In fact, only the usual vertex from the
interaction~$\bar\psi\Aslash\psi$ remains. Before we introduce a
shortcut to obtain this result directly, we have to explain how to
evaluate the path integral~(\ref{fstaction}).

Gauge invariance implies that the operator~$D_{qq}^{\mu\nu}$ given in
Eq.~(\ref{fullAprop}) has zero eigenvectors of the form
\begin{equation}
        q^L_{\mu, n} = \partial_\mu\omega_n \ .
\end{equation}
We will assume  the scalar functions~$\omega_n$ to be
eigenvectors of the d'Alembert operator, i.e.,
\begin{equation}
 - \Box \omega_n = l_n \omega_n \ .
\end{equation}
The procedure to evaluate path integrals with zero-modes is well
known~\cite{AHM,Corrigan_Goddard}.  The expansion of the
fluctuation~$q_\mu$ in terms of eigenvectors of the operator~$D_{qq}^{\mu\nu}$
is given by
\begin{equation}
        q^\mu = \sum_n a_n \xi_n^\mu + \sum_m b_m \zeta_m^\mu \ ,
\end{equation}
where~$\zeta_m^\mu = \partial^\mu\omega_m$ and~$\xi^\mu_n$ have zero
and non-zero eigenvalues, respectively.

In order to evaluate the path integral~(\ref{fstaction}), we use
Polyakov's method~\cite{Polyakov} and equip the space of fields with a
metric
\begin{eqnarray}
        ||q||^2 & = & \int{\rm d}^{\rm d}x q_\mu q^\mu \\
          & = & \sum_n a_n^2 + \sum_m b_m^2 l_m \ .
\end{eqnarray}
The volume element associated with this metric is then given by
\begin{equation} \label{measure}
        {\rm d}\mu[\kappa,\bar\kappa,q] = {\cal N} \prod_n {\rm d}a_n
\prod_m {\rm d}b_m \sqrt{\det (-\Box)} \ .
\end{equation}
The integration over the zero-modes yields the volume factor of the
gauge group, which can be absorbed by the normalisation of the
integral. The remaining integral over the non-zero modes is damped by
the usual Gaussian factor. Up to an irrelevant infinite constant one
obtains the following result for the one-loop generating functional
\begin{equation} \label{fstactionresult}
W[\eta, \bar\eta, k_{\mu\nu}]  =  \int {\rm d}^{\rm d}x\,
\lag(A^{{\rm cl}}_\mu, \psi^{{\rm cl}})
- i \ln\det\left(i\Dslash-m\right)  + {i\over2}{\ln\det}^\prime\left(-D_{qq}\right)
\end{equation}
where~${\det}^\prime (-D_{qq})$ is defined as the product of all non-zero
eigenvalues of the operator~$-D_{qq}$. In the Abelian case the
contribution from the measure, given by the determinant in
Eq.~(\ref{measure}), is trivial.  Since zero and non-zero eigenvectors
are orthogonal to each other, implying~$\partial_\mu\xi_n^\mu = 0$,
one furthermore verifies the identity
\begin{equation} 
{\det}^\prime (-D_{qq}^{\mu\nu}) = {\det\left( -D_{qq}^{\mu\nu} - \partial^\mu\partial^\nu
\right)\over \det (-\Box)} \ .
\end{equation}
Hence, up to another irrelevant infinite constant one-loop
contributions containing gauge-boson propagators are of the form
\begin{equation}
{i\over2}{\ln\det}^\prime\left(-D_{qq}\right) = 
        - {i\over2} \sum_{n=1}^{\infty}{1\over n}{\rm tr}\left( (
                        \Delta i \sigma)^n \right) \ ,
\end{equation}
with the gauge-boson propagator~$\Delta_{\mu\nu}$ given in Eq.~(\ref{GBpropagator}).

There is only one problem left. Both the fermionic determinant in
Eq.~(\ref{fstactionresult}) as well as the full fermion propagator in
Eq.~(\ref{sigmadef}) depend on the longitudinal
component of the gauge field. This can be traced back to
Eq.~(\ref{psifstfluc}). The quantity~$\kappa$ describes the
fluctuation of the gauge-independent part of the fermion field~$\psi$
as well as of its phase. To remove this dependence, one may set
\begin{equation} \label{changekappa}
\kappa(x) \rightarrow  \kappa(x) e^{ - i \int {\rm d}^{\rm d}y G_0(x-y)
\partial^\mu A^{{\rm cl}}_\mu(y) } 
\end{equation}
in Eq.~(\ref{fstaction}). As a result Eqs.~(\ref{dkk}--\ref{dqq})
involve only the fields~$\chi, \eta$ and the
derivative~$D_\mu^T$. Now the one-loop
functional~(\ref{fstactionresult}) is of the form
\begin{equation} \label{fstactionfinal}
W[\eta, \bar\eta, k_{\mu\nu}]  =  \int {\rm d}^{\rm d}x\,
\lag(A^{{\rm cl}}_\mu, \psi^{{\rm cl}})
- i \ln\det\left(i\Dslash^T-m\right)
- {i\over2} \sum_{n=1}^{\infty}{1\over n}{\rm tr}\left( (
                        \Delta i \bar\sigma)^n \right) \ ,
\end{equation}
with
\begin{equation}\label{sigmabardef}
\bar\sigma_{\mu\nu} = 2 e^2\bar\chi \gamma_\mu
(i\Dslash^T-m)^{-1} \gamma_\nu\chi \ . 
\end{equation}

\subsubsection*{More loops}

The effect of the phase in Eq.~(\ref{sndinvsource}) is understood. It
ensures that the external source~$\eta$ does not couple to the
physically unobservable phase of the fermion field~$\psi$. Its
presence, however, seems to manifest itself rather painfully during
the evaluation of the generating functional, cf.~Eqs.(\ref{Aeqmo},
\ref{dkq}-\ref{dqq}) and~(\ref{diagonaldqq}). One would expect the
problem to get even worse beyond the one-loop level.

However, this is not true. Our way to arrive at the results given in
Eqs.~(\ref{Alsol}-\ref{chisol}) and~(\ref{fstactionfinal}) was so
involved because we did not choose a convenient parametrisation for
the quantum fluctuations in Eq.~(\ref{psifstfluc}).  A better choice
is already indicated in Eq.~(\ref{changekappa}) where we had to
rescale the field~$\kappa$ in order to get rid of the gauge degree of
freedom in Eqs.~(\ref{fstactionresult}) and~(\ref{sigmadef}). Note
that we also removed the phase of the fermion field~$\psi^{{\rm cl}}$ in
Eq.~(\ref{rescalepsi}). Thus, one would better use the parametrisation
\begin{eqnarray}
A_\mu &=& A_\mu^{{\rm cl}} + \sqrt{2} e q_\mu  \ , \label{Asndfluc}\\
\psi & = & e^{-i (\phi + E)} \left({ \chi + \kappa }\right) \label{psisndfluc} \ ,
\end{eqnarray}
with
\begin{eqnarray}
\phi(x) & = &  { \int {\rm d}y G_0(x-y)
\partial^\mu A_\mu^{{\rm cl}}(y) } \ , \\
E(x)      & = & \sqrt{2} e { \int {\rm d}y G_0(x-y)
\partial^\mu q_\mu(y) } \ .
\end{eqnarray}
The transition from Eq.~(\ref{psifstfluc}) to Eq.~(\ref{psisndfluc}) can in fact be
obtained by a transformation of the variable~$\kappa$ with the
determinant of the corresponding Jacobian equal to one. One readily
verifies, that this parametrisation  directly yields 
Eqs.~(\ref{Alsol}-\ref{chisol}) and~(\ref{fstactionfinal}). The phase~$\phi$
and its variation~$E$ drop out of the source term~$\bar N \psi +
\bar\psi N$ and of the Lagrangian~$\lag_{QED}$. The same is true for the
longitudinal components of the gauge field~$A_\mu^{{\rm cl}}$ and its
fluctuation~$q_\mu$. It is now straightforward to derive the following
result for the full generating functional~(\ref{fstaction}), including
all n-loop contributions:
\begin{eqnarray} \label{sndaction}
\lefteqn{e^{i W[\eta, \bar\eta, k_{\mu\nu}]} }\\
& = &e^{i \int {\rm d}^{\rm d}x\, \lag(A_\mu^{{\rm cl}}, \psi^{{\rm cl}})}
\int d\mu[\kappa,\bar\kappa,q_\mu]
 e^{i \int {\rm d}^{\rm d}x\, 
\bar\kappa D_{\bar\kappa\kappa}\kappa +
q_\mu D_{qq}^{\mu\nu} q_\nu }  \nonumber 
\\\nonumber
& &\qquad\mbox{} \times \left\{ 1 -
  {1\over 2} \left(\int {\rm d}^{\rm d}x\, \lag^{[3]}   \right)^2 
+ {1\over 24} \left(\int {\rm d}^{\rm d}x\, \lag^{[3]} \right)^4 
- \ldots
\right\}
\ , \nonumber 
\end{eqnarray}
where
\begin{eqnarray}
D_{\bar\kappa\kappa} & = & i\Dslash^T - m \ , \label{D1q} \\
D_{qq}^{\mu\nu} & = & \Box g^{\mu\nu} -
        \partial^\mu\partial^\nu - {\rm PT}^{\mu\alpha} 
        \bar\sigma_{\alpha\beta} {\rm PT}^{\beta\nu} \ , \label{D2q}
        \\ \label{L3q} \lag^{[3]} & = & \sqrt{2} e \!\left( \bar\kappa
        + \sqrt{2} e \bar\chi\qslash^T (i\Dslash^T - m)^{-1} \right)
        \!\qslash^T\!\!\left( \kappa + \sqrt{2} e (i\Dslash^T -
        m)^{-1} \qslash^T\chi \right),\ \ \ \ \ 
\end{eqnarray}
and
\begin{equation}
  q^T_\mu = {\rm PT}_{\mu\nu} q^\nu \ .
\end{equation}
Any vertex emits non-zero modes only. The full gauge field propagator
entering loop graphs is given by the inverse of the
operator~$D_{qq}^{\mu\nu}$, restricted to the subspace of
non-zero modes. Thus, the evaluation of loop contributions of arbitrary
order boils down to the evaluation of Gaussian integrals, for
example, of the form
\begin{equation}
  2 \int {\rm d}\mu[q] q_\mu^T(x) q_\nu^T(y) e^{i\int{\rm d}^{\rm d}x
    q_\alpha D_{qq}^{\alpha\beta} q_\beta} = i \langle x |
  {-D_{qq}^{-1} }^\prime_{\mu\nu} | y \rangle / \sqrt{
    {\det}^\prime\left( -D_{qq}\right) }\ ,
\end{equation}
where
\begin{equation}
  {D_{qq}^{-1} }^\prime_{\mu\rho} D_{qq}^{\rho\nu} = D_{qq}^{\nu\rho}
  {D_{qq}^{-1} }^\prime_{\rho\mu} = {\rm PT}^\nu_\mu \ .
\end{equation}
One readily verifies that
\begin{eqnarray}\nonumber
  -i {D_{qq}^{-1} }^\prime & = & -i {\rm PT}\left( \Box - {\rm
    PT}\bar\sigma{\rm PT} \right)^{-1} {\rm PT} \\ & = &
  \sum_{n=0}^\infty\left(\Delta i \bar\sigma\right)^n \Delta \ ,
  \label{finalfullprop}
\end{eqnarray}
with the free gauge field propagator~$\Delta_{\mu\nu}$ given in
Eq.~(\ref{GBpropagator}). The corresponding full fermion propagator is
\begin{equation}
  -i \left(i \Dslash^T - m\right)^{-1} = \sum_{n=0}^\infty\left(S i
  \Aslash^{T, {\rm cl}}\right)^n S ,
\end{equation}
with the free fermion propagator~$S$ given in Eq.~(\ref{Spropagator}).

\subsection{QED the old-fashioned way}

The following discussion of the Faddeev-Popov ansatz~(\ref{fadpopans})
for the generating functional $W_f[\eta, \bar\eta, k_{\mu\nu}]$ is
based on the gauge condition
\begin{equation} \label{expgaugefunc}
  f(A_\mu) = \partial_\mu A^\mu \ .
\end{equation}
In this case the determinant of the matrix $M_f$ yields an infinite
constant which can be absorbed by the normalisation of the path
integral. The delta function can be converted into an exponential
factor. One obtains the well known result
\begin{equation} \label{fadpopaction}
  e^{i W_f^\xi[\eta, \bar\eta, k_{\mu\nu}]} = \int
  d\mu[\psi,\bar\psi,A_\mu] e^{i \int {\rm d}^{\rm d}x\, \lag_{QED} +
    \lag_{GF} + \bar\eta \psi + \bar\psi\eta + k^{\mu\nu} F_{\mu\nu}}
  \ ,
\end{equation}
with
\begin{equation} \label{expgaugecond}
  \lag_{GF} = {-1\over 2 \xi e^2}\left(\partial_\mu A^\mu\right)^2 \ .
\end{equation}

Since it is straightforward to repeat the whole analysis of the last
subsection for the case at hand we merely state the results and
compare.
The classical fields satisfy the equations of motion
\begin{eqnarray}
  \left(i\Dslash - m\right)\psi^{{\rm cl}} &=& -\eta \\ 
    \label{Aeqmofixed}
\partial^\mu F_{\mu\nu}^{{\rm cl}} + {1\over\xi}\partial^\mu\partial_\nu
A_\mu^{{\rm cl}} &=& 2 e^2 \partial^\mu k_{\mu\nu} + e^2
\bar\psi^{{\rm cl}}\gamma_\nu\psi^{{\rm cl}}
\end{eqnarray}
to be compared with Eqs.~(\ref{Neqmo}) and~(\ref{AAeqmo}). In this
case the electromagnetic current is not conserved either. Here however
its divergence does enter the equations of motion. It determines the
gauge degree of freedom. Due to the presence of the gauge-dependent
piece on the left hand side of Eq.~(\ref{Aeqmofixed})
problem~(\ref{condition}) does not occur here.
The solutions are
\begin{eqnarray}
  A^{{\rm cl}, \xi}_\mu(x) & = & i e^2 \int d^dy
  \Delta^\xi_{\mu\nu}(x-y) \left( 2 \partial_\rho k^{\rho\nu}(y) +
  \left(\bar\psi^{{\rm cl}, \xi}\gamma^\nu\psi^{{\rm cl},
      \xi}\right)(y)\right) \ , \\ \psi^{{\rm cl}, \xi}(x) &=& i \int
    {\rm d}^{\rm d}y S(x-y) \left( -\eta(y) + \left( \Aslash^{{\rm
        cl}, \xi}\psi^{{\rm cl}, \xi}\right)(y)\right) \ ,
\end{eqnarray}
corresponding to Eqs.~(\ref{Alsol}-\ref{atsol}) and~(\ref{chisol}).
They determine the gauge degree of freedom as well. The gauge-field
propagator $\Delta^\xi_{\mu\nu}$ is of the form
\begin{equation}\label{GDGBpropagator}
  \Delta^\xi_{\mu\nu}(x-y) = i \langle x| {{\rm PT}_{\mu\nu}\over
    -\Box} + \xi {{\rm PL}_{\mu\nu}\over -\Box} |y\rangle \ ,
\end{equation}
corresponding to Eq.~(\ref{GBpropagator}). The fermion propagator is
given in Eq.~(\ref{Spropagator}). In the limit~$\xi\rightarrow0$
(Landau gauge) one obviously recovers the results given in
Eqs.~(\ref{Alsol}-\ref{atsol}) and~(\ref{chisol}) for the special
case~$\omega=0$. In this case, the longitudinal component of the gauge
field vanishes.

Contributions from loops work out in the same way. Since the gauge is
fixed, zero-modes do not occur and the evaluation of the path integral
corresponding to Eq.~(\ref{sndaction}) is straightforward. Instead of
the quantities in Eqs.~(\ref{D1q}, \ref{D2q}) and~(\ref{L3q}) it involves
\begin{eqnarray}
  D_{\bar\kappa\kappa}^\xi & = & i\Dslash - m \ , \\ D_{qq}^{\xi,
    \mu\nu} & = & \Box g^{\mu\nu} -
  \left(1-{1\over\xi}\right)\partial^\mu\partial^\nu -
    \sigma_{\alpha\beta} \ , \\ \lag^{\xi[3]} & = & \sqrt{2} e\!
    \left( \bar\kappa + \sqrt{2} e\bar\psi^{{\rm cl}, \xi}\qslash
      (i\Dslash - m)^{-1} \right)\!\qslash\!\left( \kappa + \sqrt{2}
      e (i\Dslash - m)^{-1} \qslash\psi^{{\rm cl}, \xi} \right) . \ \ 
      \ 
\end{eqnarray}
The one-loop contribution corresponding to the result in
Eq.~(\ref{fstactionfinal}) is of the form
\begin{equation}
- i \ln\det\left(i\Dslash-m\right)
- {i\over2} \sum_{n=1}^{\infty}{1\over n}{\rm tr}\left( (
                        \Delta^\xi i \sigma)^n \right) \ .
\end{equation}
In Landau gauge both are the same. 

Contributions beyond one-loop order involve the full gauge field
propagator~$-i {D_{qq}^\xi}^{-1}$.  In the limit~$\xi\rightarrow0$ it
reduces to Eq.~(\ref{finalfullprop}). Hence, in Landau gauge longitudinal
fluctuations of the gauge field do not contribute either.

\subsection{Feynman rules}

The Feynman rules to evaluate the gauge-invariant generating
functional $W[\eta, \bar\eta, k_{\mu\nu}]$ given in
Eq.~(\ref{pathrep}) are exactly the same as those to evaluate the
gauge-dependent generating functional~$W_f^{\xi=0}[\eta, \bar\eta,
k_{\mu\nu}]$ defined in Eq.~(\ref{fadpopaction}) in Landau gauge.
Corresponding Green functions are identical.

This includes a photon mass which might be introduced as an infrared
regulator before some of the external momenta are set on the mass-shell. The
mass term
\begin{equation}
        {1\over 2} \mu^2 A_\mu A^\mu
\end{equation}
which is usually introduced in Eq.~(\ref{fadpopaction}) is not gauge
invariant. Hence, in order to preserve gauge symmetry, one has to
use the non-local mass term
\begin{equation}
        {1\over 2} \mu^2 A_\mu^T A^{T\mu}
\end{equation}
in Eq.~(\ref{pathrep}) instead, which involves  only the transverse components
of the gauge field. Again one obtains the same Feynman rules.

Note that off-shell momenta already provide an infrared
regulator. Thus, our analysis presented above was not obstructed by
infrared problems, since it was concerned with Green's functions only.

\section{Discussion}

In the following discussion of QED we will adopt a language that
equally applies to the non-Abelian case. Thus, we will talk of zero-
and nonzero-modes, instead of longitudinal and transversal components.

There are two approaches to define Green's functions for charged
particles. One involves a non-local source of the form given in
Eq.~(\ref{sndinvsource}) which defines gauge-invariant Green functions. The
other involves gauge-dependent sources and is based on the
Faddeev-Popov ansatz in Eq.~(\ref{fadpopans}); in this case Green's functions
depend on both, the gauge condition and the gauge-fixing
parameter~$\xi$. Surprisingly, both approaches yield exactly the same
Green functions in Landau gauge, i.e., for the gauge
condition~(\ref{expgaugefunc}) and in the limit~$\xi\rightarrow0$.

The quadratic form in the exponent of Eq.~(\ref{sndaction}) has
zero-modes which is a consequence of gauge invariance. These
zero-modes correspond to fluctuations of the fields that are gauge
transformations. This divides the space of all fluctuations into two
parts. The subspace of all nonzero-modes is the physical one. The
essential feature of the manifestly gauge-invariant approach is that
it is restricted to this subspace only. In the equations of
motion~(\ref{Neqmo}) and~(\ref{AAeqmo}) the gauge degree of freedom
drops out. One-loop corrections involve the determinant of a
differential operator defined on the subspace of nonzero-modes. The
inverse of the same operator, i.e., the full propagator, is also
defined on this subspace. In Eq.~(\ref{sndaction}) the integration
over the zero-modes yields the volume factor of the gauge group and is
absorbed by the normalisation of the integral. This explains why the
exponential factor of the source in Eq.~(\ref{sndinvsource}) does not
produce any vertex. It involves only the longitudinal component of the
gauge field which does not affect the dynamics of the theory. The
other source in Eq.~(\ref{fstinvsource}) is a different matter. It
also involves the physical components of the gauge field.

If a gauge-fixing condition is introduced, like in
Eq.~(\ref{fadpopans}), all quantities are extended to the full space.
All modes propagate (see the propagator in
Eq.~(\ref{GDGBpropagator})). In the case of gauge-invariant sources,
the unphysical contributions from the subspace of zero-modes are
exactly cancelled order by order by the corresponding contributions
from ghosts, which are trivial in the Abelian case. This cancellation
does not happen if the sources couple to the gauge degree of freedom;
the resulting Green functions are gauge-dependent.

In the limit~$\xi\rightarrow0$ the exponential factor in
Eq.~(\ref{fadpopaction}) involving the gauge-fixing part of the Lagrangian
becomes a delta-function which confines the longitudinal component of
the gauge field and its fluctuation to zero. This confines the whole
analysis based on this ansatz to the subspace of transversal modes.
Furthermore, under this constraint, there is no difference between the
two sources~$N$ and~$\eta$. This explains why the two approaches yield
the same Green functions and the same Feynman rules in this limit.

This result also proves that the exponential factor in
Eq.~(\ref{sndinvsource}) does neither spoil the renormalizability of
the theory nor affect its infrared properties. From a physical point
of view this was to be expected. The phase does only involve the
longitudinal component of the gauge field which does not affect the
dynamics. Renormalization is in fact simpler in the manifestly gauge
invariant approach. Since the symmetry is manifest at any step, all
counter terms must necessarily be gauge invariant. Thus, for example,
the wave function renormalization of any gauge field, Abelian or
non-Abelian, must be the inverse of the corresponding coupling
constant renormalization. One does not need to employ Ward identities
to establish such relations.

\section{The Standard Model}

Gauge-invariant sources are readily constructed for all particles in
the Standard Model of electroweak interactions. First we observe
that the complex Higgs-doublet field~$\phi$ and its conjugate field
\begin{equation}
\tilde\phi = i \tau^2 \phi^*
\end{equation}
provide a natural reference frame for~$SU_L(2)$-valued fields~\cite{tHooft}. In what follows
we will furthermore use the decomposition
\begin{equation}
\phi = {m\over\sqrt{\lambda}} R U \ ,
\end{equation}
where the unitary field~$U$, satisfying~$U^\dagger U = 1$, describes the three
Goldstone bosons while the radial component~$R$ represents the Higgs boson.
 
For up and down type fermions one may consider the composite fields
\begin{equation}\begin{array}{rclrcl}
\phi^\dagger \Psi^i_L & = & {m\over\sqrt{\lambda}} R d_L^i \ , \qquad &
\tilde \phi^\dagger \Psi^i_L & = &  {m\over\sqrt{\lambda}} R u_L^i  \ , \\
\phi^\dagger \Psi^i_L & = & {m\over\sqrt{\lambda}} R e_L^i \ , \qquad &
\tilde \phi^\dagger \Psi^i_L & = &  {m\over\sqrt{\lambda}} R \nu_L^i \ , 
\end{array}\label{fcomposite}
\end{equation}
where the interpolating fermion fields
\begin{equation}\begin{array}{rclrcl}
d_L^i & = & U^\dagger \Psi^i_L \ , \qquad &
u_L^i  & = &  \tilde U^\dagger \Psi^i_L \ , \\
e_L^i & = & U^\dagger \Psi^i_L \ , \qquad &
\nu_L^i  & = &  \tilde U^\dagger \Psi^i_L\ ,
\end{array}\label{ffields}
\end{equation}
are~$SU_L(2)$-invariant and have the same quantum numbers as their right-handed
counterparts. The quantities~$\Psi_L^i$ are the left-handed iso-doublet
fields. For the massive gauge bosons one may consider the composite fields
\begin{equation}\begin{array}{rcrcl}
V_\mu^1 & = & i \tilde\phi^\dagger D_\mu\phi + i \phi^\dagger D_\mu\tilde\phi 
& = & {m2\over\lambda} R^2 {\cal W}_\mu^1 \\
V_\mu^2 & = & - \tilde\phi^\dagger D_\mu\phi +  \phi^\dagger D_\mu\tilde\phi 
& = & {m2\over\lambda} R^2 {\cal W}_\mu^2 \\
V_\mu^3 & = & i \tilde\phi^\dagger D_\mu\tilde\phi - i \phi^\dagger D_\mu\phi 
& = & {m2\over\lambda} R^2 {\cal Z}_\mu
\end{array}\label{bcomposite}
\end{equation}
where the gauge boson fields
\begin{eqnarray}\label{bfieldsf}
{\cal W}^+_\mu &=& {i\over2} \left(\tilde U^\dagger (D_\mu U) - (D_\mu\tilde U)^\dagger U\right) \\
{\cal W}^-_\mu &=& {i\over2} \left( U^\dagger (D_\mu \tilde U) - (D_\mu U)^\dagger \tilde U\right) \\
{\cal Z}_\mu &=& i \left( \tilde U^\dagger (D_\mu \tilde U) - U^\dagger (D_\mu U)\right) \\
{\cal A}_\mu &=& B_\mu + s^2_W {\cal Z}_\mu \label{bfieldsl} \\
{\cal W}^\pm_\mu &=& {1\over2} ({\cal W}^1_\mu \mp i {\cal W}^2_\mu) 
\end{eqnarray}
are also~$SU_L(2)$-invariant. The covariant derivative
\begin{equation}
D_\mu = \partial_\mu - {i\over2} \tau^a W^a_\mu - {i\over2} Y B_\mu 
\end{equation}
involves the~$U_Y(1)$ gauge field~$B_\mu$ and the~$SU_L(2)$
triplet~$W_\mu^a$. Up to a constant factor the composite
fields~$V_\mu^i$ in Eq.~(\ref{bcomposite}) correspond to the currents
of the custodial symmetry~$SU_c(2)$. The gauge-boson fields in
Eqs.~(\ref{bfieldsf}-\ref{bfieldsl}) are long known from the effective
field theory analysis of the symmetry-breaking sector of the Standard
Model~\cite{Georgi}.

External sources can be coupled to the~$SU_L(2)$-invariant composite
fields in Eqs.~(\ref{fcomposite}) and~(\ref{bcomposite}). In order to make them invariant
under the full group~$SU_L(2)\times U_Y(1)$ one may introduce
phase factors for all charged par\-ti\-cles in the same fashion as in
Eq.~(\ref{sndinvsource}), involving the Abelian gauge field~$B_\mu$.
One may choose
\begin{eqnarray}\label{SMsources}
\lefteqn{\lag_{source} = {1\over2} h \phi^\dagger\phi + k_{\mu\nu} B^{\mu\nu} + 
J^{a\mu} V^a_\mu  } \\
& & \mbox{} + \bar N_L^i \phi^\dagger\Psi_L^i + \bar \Psi_L^i \phi N_L^i +
\bar M_L^i \tilde\phi^\dagger\Psi_L^i + \bar \Psi_L^i \tilde\phi M_L^i +
\bar N_R^i \psi_R^i + \bar \psi_R^i N_R^i \ , \nonumber
\end{eqnarray}
with external fields~$N_L^i, M_L^i, N_R^i, J^a_\mu, k_{\mu\nu}$
and~$h$. One has,  for example,
\begin{equation}
 J_\mu(x) = j_\mu(x) e^{ - T \int {\rm d}^{\rm d}y G_0(x-y)
\partial_\mu B^\mu(y) } 
\end{equation}
with
\begin{equation}
T = \left(
\begin{array}{rrr}
0 & 1 & 0 \\
-1 & 0 & 0 \\
0 & 0 & 0
\end{array}\right) \ .
\end{equation}
The sources in Eq.~(\ref{SMsources}) define~$SU_L(2)\times
U_Y(1)$-invariant Green functions of the fields in Eqs.~(\ref{ffields})
and~(\ref{bfieldsf}-\ref{bfieldsl}), of the Higgs field~$R$ and of the right-handed fermion
fields~$\psi_R^i$. Note that the vacuum in the spontaneously broken
phase corresponds to the value~$R=1$. The
occurrence of the Higgs-boson field in the sources of the left-handed fermions
and of the gauge bosons will give rise to external line
renormalizations, which are under control. The effect of the phase was
thoroughly investigated in Section~2.

To evaluate these Green functions perturbatively one may employ the
gauge-invariant technique of Section~2, which is not restricted to the
Abelian case~\cite{Corrigan_Goddard}.

In order to analyse how Green's functions defined through the sources in
Eq.~(\ref{SMsources}) relate to the familiar gauge-dependent ones we define
the Faddeev-Popov ansatz with the gauge-fixing condition
\begin{equation}
\lag_{GF} = {-1\over2\xi_B {g^\prime}^2} \left(\partial_\mu B^\mu\right)^2
+ {-1\over2\xi_Wg^2}\left(\partial^\mu W^a_\mu 
 - i \xi_W {m g^2 \over\sqrt{\lambda}}
 \left( U_0^\dagger\tau^a\phi-\phi^\dagger\tau^aU_0\right)\right)^2 \ ,
\end{equation}
and the unitary-gauge version of the sources given above, i.e., with
\begin{equation}
U \rightarrow U_0 = \left(
\begin{array}{c}
0 \\ 1
\end{array}
\right) 
\end{equation}
set in the source terms.  The phase factors involving the longitudinal
component of the $U_Y(1)$ gauge field~$B_\mu$ are also dropped. Apart
from external line renormalizations due to the presence of the Higgs
field~$R$ this defines gauge-dependent Green functions commonly used
in the literature.

As far as the~$U_Y(1)$ symmetry is concerned we have exactly the same
situation as in the theory of QED. The non-Abelian part of the
electroweak group is a different matter. The source terms of our
Faddeev-Popov ansatz for the Standard Model were defined as the
unitary-gauge limit of the corresponding~$SU_L(2)\times U_Y(1)$
invariant source terms. Hence, we expect Green's functions of this
Faddeev-Popov ansatz to be the same as those of the manifestly
gauge-invariant framework if the limits~$\xi_B\rightarrow 0$ (Landau
gauge with respect to the group~$U_Y(1)$) and~$\xi_W
\rightarrow\infty$ (unitary gauge with respect to the group~$SU_L(2)$)
are taken.  It is indeed straightforward to verify this statement at
tree-level. We did not go beyond that. 

Since a gauge-dependent calculation in unitary gauge is plagued by the
bad ultra-violet behaviour of the gauge-boson propagators it is not
clear how this statement generalises to loop corrections. However, one
should be very careful not to draw any wrong conclusion from this
relation between gauge-invariant and gauge-dependent Green
functions at tree-level. The bad ultra-violet behaviour is a specific
problem of the gauge-dependent approach in unitary gauge. It does not
occur in the manifestly gauge-invariant approach we described. We
particularly emphasise that loop corrections in the gauge-invariant
framework have a decent ultra-violet behaviour~\cite{AHM}. For large
momenta~$k$ all propagators fall off as~$k^{-2}$.

We conclude this section with some remarks about renormalization.
Dimensional arguments show that the source terms in
Eq.~(\ref{SMsources}) are renormalizable. The phase does not affect
this property as shown in Section~2. Green's functions of the
composite operators in Eq.~(\ref{fcomposite}) and~(\ref{bcomposite}),
however, are more singular at short distances than (gauge-dependent)
Green functions of the fields~$\phi, \Psi_L^i, B_\mu, \ldots$. Time
ordering of these operators gives rise to ambiguities and
corresponding Green functions are only unique up to contact terms.
In order to make the theory finite, these contact terms of dimension
four need to be added to the Lagrangian. A detailed discussion of this
point is deferred to an explicit analysis of the Standard Model within
this framework~\cite{inpreparation}. As for the Abelian Higgs model
this is already discussed in Ref.~\cite{AHM}.

\section{Summary}

This work provides a manifestly gauge-invariant approach to analyse
gauge theories with charged particles. Its two main ingredients are
(1) Green functions of gauge-invariant operators and (2) Feynman
rules which do not involve any kind of gauge fixing.

In the first part of this article we thoroughly discussed the theory
of QED.  We considered Green functions of the fermion field~$\psi$ and of
the electromagnetic field strength~$F_{\mu\nu}$ which are generated by the 
vacuum-to-vacuum transition amplitude
\begin{equation} \label{vacvacs}
\langle 0_{\rm out}|0_{\rm in}\rangle_{N,\bar N, k_{\mu\nu}} 
=\langle0| T\left[ e^{i \bar N \psi_{\rm op} + i \bar\psi_{\rm op} N
+ i k_{\mu\nu} F^{\mu\nu}_{\rm op}} \right]|0\rangle \ .
\end{equation}
In order to generate gauge-invariant Green functions the external
sources must not couple to the gauge degree of freedom.  Hence, the
external source of the fermion, $N$, must be gauge-variant, with the
same transformation properties as the fermion field itself.  Moreover,
the vacuum-to-vacuum transition amplitude would not even be defined,
if the external field~$N$ were a fixed function of space-time.

A very convenient source with the afore mentioned property is of the
form
\begin{equation} \label{sndinvsources}
N(x) = \eta(x) e^{ - i \int {\rm d}^{\rm d}y G_0(x-y)
\partial_\mu A^\mu(y) } \ ,
\end{equation}
where~$\eta$ is a singlet under the gauge-group. The free massless
propagator~$G_0(z)$ is defined in Eq.~(\ref{masslessprop}). With this
source the amplitude in Eq.~(\ref{vacvacs}) generates gauge-invariant
Green functions of the fermion field and of the electromagnetic field
strength which are obtained by taking derivatives with respect to the
external fields~$\eta$ and~$k_{\mu\nu}$.

The phase in Eq.~(\ref{sndinvsources}) does only involve the
longitudinal component of the gauge field which does not affect the
dynamics of the theory. Hence it does not produce any
vertex. Furthermore, it can neither harm renormalizability
nor affect the infrared behaviour of the theory. It has only one
effect, to ensure that the extenal field~$\eta$ does not couple to the
gauge degree of freedom.

We showed how the path-integral representation of the vacuum-to-vacuum
transition amplitude in Eq.~(\ref{vacvacs}) can be evaluated
perturbatively without fixing the gauge. The technique involves only
the physical degrees of freedom. The gauge degrees of freedom, i.e.,
the longitudinal component of the gauge field, the phase of the
fermion field, as well as their quantum fluctuations drop out
completely. As a result, the propagator of the gauge field is only
defined on the subspace of transversal modes. The application of this
manifestly gauge-invariant technique is necessary in order to avoid
the problems discussed in the introduction.

A remarkable result is that the gauge-invariant Green functions
defined through Eqs.~(\ref{vacvacs}) and~(\ref{sndinvsources}) are
exactly the same as the gauge-dependent Green functions defined
through the Faddeev-Popov ansatz in Eq.~(\ref{fadpopaction}) in Landau
gauge. Feynman rules in both approaches are also the same. This is
readily understood. In the gauge-dependent framework the gauge-field
propagator is defined on the full space, involving transversal and
longitudinal components. In Landau gauge the longitudinal components
are switched off and all quantities are confined to the physical
subspace of transversal modes.

In the second part of this article we showed how the gauge-invariant
approach can be extended to the Standard Model of electroweak
interactions. The Higgs-doublet field~$\phi$ and its conjugate field
provide a frame of reference for~$SU_L(2)$-valued quantities. Since
the symmetry is spontaneously broken this property can be used to
define~$SU_L(2)$-invariant fields for fermions and bosons. For neutral
particles these fields are already invariant under the full
electroweak group~$SU_L(2)\times U_Y(1)$.  For charged particles, on
the other hand, one may consider external sources similar to the one
given in Eq.~(\ref{sndinvsources}) involving the longitudinal
component of the~$U_Y(1)$ gauge field~$B_\mu$. This provides
~$SU_L(2)\times U_Y(1)$-invariant sources for all particles of the
theory.  The technique to evaluate the corresponding gauge-invariant
Green functions without gauge fixing is the same as for the Abelian
case.  For details on the Standard Model, in particular for the
comparison of the gauge-invariant approach to the Faddeev-Popov
ansatz, the reader is refered to Section~4.

We have not discussed whether the gauge-invariant approach described
in this article can also be applied to the theory of QCD. The open
question is whether one can construct colour-singlet operators which
create single quarks or gluons. To do so one might, for example,
attempt to generalise the source term given in
Eq.~(\ref{sndinvsources}). However, since colour is confined it may as
well turn out that all physically relevant information can be
extracted from Green's functions of operators that create pairs of
these particles.

\section*{Acknowledgements}

The author is particularly indebted to G.\ Weiglein for information on
the BGFM and on gauge-fixing.  Furthermore, it is a pleasure to him to
thank J.~Gasser, J.\ K\"uhn, H.~Leutwyler, and A.\ Nyffeler for interesting
discussions and for encouragement to write this article.

\newcommand{\artref}[4]{{\rm #1}, {\it #2} {\bf #3} #4}
\newcommand{\cartref}[5]{{\rm #1}, {\it #2},  #3 {\bf #4}, #5}
\newcommand{\bookref}[2]{{\rm #1}, #2}

\end{document}